\pdfoutput=1

\documentclass{Arxive}
\usepackage{graphicx}
\usepackage{epstopdf, epsfig}
\usepackage{amsmath}

\allowdisplaybreaks[1]

\shorttitle{Theoretical analyses of filtered Eulerian drag force}
\shortauthor{X. Chen, M. Jiang and Q. Zhou}

\title{Theoretical analyses of the sub-grid quantities' effect on filtered Eulerian drag force}

\author{Xiao Chen\aff{2},
  Ming Jiang\aff{2}
 \and Qiang Zhou\aff{1,2}
	\corresp{\email{zhou.590@mail.xjtu.edu.cn}}}

\affiliation{\aff{1}State Key Laboratory of Multiphase Flow in Power Engineering, Xi’an Jiaotong University, Xi’an 710049, China
\aff{2}School of Chemical Engineering and Technology, Xi’an Jiaotong University, Xi’an 710049, China}

\begin{document}

\maketitle

\begin{abstract}
An expression of the filtered Eulerian drag force is proposed based on the second order Taylor polynomial approximation of the microscopic Eulerian drag coefficient. Theoretical computations of the expression are performed at low Reynolds numbers based on an available microscopic drag model. It shows that four sub-grid quantities, i.e., the gas drift velocity, the solid drift velocity, the scalar variance of solid volume fraction and a third-order moment, defined as the covariance of squared solid volume fraction and the slip velocity, are significant for an accurate estimation of the filtered drag force at low Reynolds numbers. The gas drift velocity is nothing but the drift velocity defined by Parmentier et al. (AIChE Journal, 2012, 58 (4): 1084-1098), and in analogy to the gas drift velocity, we defines the solid drift velocity in the present work. The great relevance of the gas drift velocity and the scalar variance of solid volume fraction to the filtered drag force have been demonstrated through numerous correlative analyses of fully resolved simulations. Present theoretical analysis shows that the gas drift velocity term is primarily responsible for the drag reduction, whereas the solid drift velocity term is prone to attenuate this reduction, especially at moderate and high solid volume fractions. The scalar variance of solid volume fraction term is found to increase the filtered drag force in the full range of solid volume fractions. The third-order term has exactly the same coefficient with that of the variance term, and it mostly attains negative values with the trend to decrease the filtered drag force. 
\end{abstract}


\section{Introduction}
Gas-solid flows in industrial-scale fluidized beds contain trillions of particles with complicated meso-scale flow structures \citep{Agrawal2001,Sundaresan2018,Li2018}. The details of the structures are often unaffordably expensive to resolve even in the efficient two-fluid model (TFM, also known as Euler-Euler model) simulations \citep{Gidaspow1994}. Thus, to overcome this problem, one of the most promising approach, called filtered TFM, is being developed to simulate large-scale flow problems \citep{Agrawal2001,Igci2011a,Ozel2013,Parmentier2012,Schnei2017}. In filtered TFM, only the filtered transport equations on the “coarse-grid” are solved. The effects of unresolved structures at the sub-grid scales are accounted for through closures.

Among all the constitutive terms of filtered TFM’s equations, the filtered drag term is the most important one \citep{Ozel2013,Parmentier2012}. Earlier researchers obtained the filtered drag force simply using filtered quantities that are explicitly available in filtered TFM simulations. \citet{Igci2011a} firstly proposed a model for the filtered drag force as a function of the filtered volume fraction and the filter size. Then \citet{Parmentier2012} and \citet{Sarkar2016} included the filtered slip velocity in estimating the filtered drag force. However, recently works \citep{Ozel2017,Rubinstein2017,Schnei2017,Sundaresan2018} suggested that the use of only coarse grid resolved quantities is not sufficient to predict the filtered drag force since the effect of inhomogeneous sub-grid structures is not accounted for. It is found that the drag correction due to the sub-grid structures could be expressed in terms of sub-grid quantities that are not directly available in filtered TFM simulations. The drift velocity, defined as the difference between the filtered gas velocity seen by the particle phase and the resolved filtered gas velocity, was first proposed in predicting filtered drag force by \citet{Parmentier2012}. Then \citet{Ozel2013} demonstrated that the drift velocity could help produce better estimation of solid flux in a 3D dilute periodic channel flow. This quantity is referred to as the gas drift velocity hereafter. In analogy to the gas drift velocity, the solid drift velocity is also defined in the present work. Recently, \citet{Schnei2017} applied a first order Taylor series expansion on the filtered drag force and suggested that the filtered drag force is highly relative to the correlations between solid volume fraction fluctuation and Favre fluctuating velocities of the gas and solid phases. In present work, we show that these correlations could be mathematically represented by gas and solid drift velocities. Furthermore, \citet{Ozel2017} showed that the scalar variance of the solid volume fraction, which is a measure of the degree of inhomogeneous distribution of particles within a filtered volume, as a second sub-grid marker in estimating the filtered drag force. These works are consistent with Rubinstein’s direct numerical simulations in much smaller systems \citep{Rubinstein2017} and TFM simulations in much larger systems \citep{Ozel2013,Parmentier2012}. However, theoretical analyses of the specific influence of the mentioned sub-grid quantities, i.e., the drift velocities and the scalar variance of solid volume fraction, are still lacking. Also, one important question remains: is there any other sub-grid quantity playing a significant role in accurately predicting the filtered drag force?

In this paper, we use the second order Taylor expansion of the microscopic Eulerian drag coefficient to approximate the filtered Eulerian drag force. Various sub-grid quantities appear in the approximation. The contribution of them to the filtered drag force is analyzed through theoretical computation. It is noted that the scalar variance of solid volume fraction that has been shown to correlate well with the filtered drag comes into play only when the expansion is made to second order. Then the new expression of filtered Eulerian drag force is verified using the data in the literature.  
\vspace{-20pt}
\section{Filtered procedure and sub-grid quantities}
The derivation of the filtered model equations and the residual correlations requires a box filtering technique \citep{Rubinstein2017,Sarkar2016}, and the filtered part of variable $\Psi$ could be written as

\begin{equation}\label{eq1}
	\bar{\Psi}=\iiint \Psi G(x,y,z)\mathrm{d}\bold{r}
\end{equation}
where G(x,y,z) is a weight function which has the property $\iiint G(x,y,z)\mathrm{d}\bold{r}=1$. The overbar on the top of a variable in Eq.(\ref{eq1}) denotes the filtered value at the coarse grid. The fluctuating quantity ${\Psi}'$ is defined as the difference between the local quantity $\Psi$ and the corresponding filtered value $\bar{\Psi}$ at the coarse grid where the local quantity belongs to.
\begin{equation}\label{eq3}
	{\Psi}'=\Psi-\bar{\Psi}
\end{equation}
with the above definition, the identity $\overline{\bar{\Psi}}=\bar{\Psi}$ could be simply obtained. Further using this box filter, Favre-averaged flow variables and their corresponding fluctuating values could be defined as:
\begin{equation}\label{eq4}
	\tilde{\Psi}=\frac{\overline{\phi_k\Psi_k}}{\bar{\phi}_k}
\end{equation}
\begin{equation}\label{eq5}
	{\Psi}''=\Psi_k-\tilde{\Psi}_k
\end{equation}
where the subscript $k$ denotes the different phases ($k=s, g$). Following the studies by \citet{Parmentier2012} and \citet{Ozel2013}, the sub-grid drift velocity is defined as
\begin{equation}\label{eq6}
	\bar{\phi}_s\tilde{v}_{d,i}=\overline{\phi_s(u_{g,i}-u_{s,i})}-\bar{\phi}_s(\tilde{u}_{g,i}-\tilde{u}_{s,i})=\overline{\phi_su_{g,i}}-\bar{\phi}_s\tilde{u}_{g,i}
\end{equation}
where $\tilde{v}_{d,i}$ is the drift velocity in {\it i} direction; $u_{g,i}$ and $u_{s,i}$ are the gas and solid velocities in {\it i} direction. Take the volume-average of Eq. (\ref{eq5}), then combine with the identity $\bar{u}_{g,i}+{u}'_{g,i}=\tilde{u}_{g,i}+{u}''_{g,i}$ yielding $\overline{{u}''_{g,i}}={u}''_{g,i}-{u}'_{g,i}$. Considering $\phi_s=1-\phi_g$, we have
\begin{eqnarray}\label{eq7}
	\bar{\phi}_s\tilde{v}_{d,i} & = & \bar{u}_{g,i}-\overline{\phi_gu_{g,i}}-\bar{\phi}_s\tilde{u}_{g,i}=\bar{u}_{g,i}-\tilde{u}_{g,i}=\bar{u}_{g,i}-\frac{\overline{\phi_gu_{g,i}}}{\bar{\phi}_g}\nonumber\\
& = & \bar{u}_{g,i}-\frac{\bar{\phi}_g\bar{u}_{g,i}+\overline{{\phi}'_g{u}'_{g,i}}}{\bar{\phi}_g}=-\frac{\overline{{\phi}'_g{u}'_{g,i}}}{\bar{\phi}_g}=-\frac{\overline{{\phi}'_g({u}''_{g,i}-\overline{{u}''_{g,i}})}}{\bar{\phi}_g}=-\frac{\overline{{\phi}'_g{u}''_{g,i}}}{\bar{\phi}_g}
\end{eqnarray}
Similarly, we define the solid drift velocity and summarize the gas and solid drift velocities as follows:
\begin{equation}\label{eq10}
	\tilde{v}_{d,g,i}=\frac{\bar{u}_{g,i}-\tilde{u}_{g,i}}{\bar{\phi_s}}=\frac{\overline{{u}''_{g,i}}}{\bar{\phi_s}}=-\frac{\overline{{\phi}'_g{u}'_{g,i}}}{\bar{\phi_s}\bar{\phi}_g}=-\frac{\overline{{\phi}'_g{u}''_{g,i}}}{\bar{\phi_s}\bar{\phi}_g}
\end{equation}
\begin{equation}\label{eq11}
	\tilde{v}_{d,s,i}=\frac{\bar{u}_{s,i}-\tilde{u}_{s,i}}{\bar{\phi_g}}=\frac{\overline{{u}''_{s,i}}}{\bar{\phi_g}}=-\frac{\overline{{\phi}'_s{u}'_{s,i}}}{\bar{\phi}_s\bar{\phi_g}}=-\frac{\overline{{\phi}'_s{u}''_{s,i}}}{\bar{\phi}_s\bar{\phi_g}}
\end{equation}
According to Eqs. (\ref{eq10}) and (\ref{eq11}), Scheiderbauer’s conclusion\citep{Schnei2017,Schnei2018} that the filtered drag force is highly relative to the covariances $\overline{{\phi}'_s{u}''_{g,i}}$ and  $\overline{{\phi}'_s{u}''_{s,i}}$, is indeed the demonstration of important relevance between the filtered drag force and drift velocities for the gas and solid phases.
 
\section{Filtered drag force}
Since the interaction between the fluid and particle phases cannot be fully resolved in coarse-grid simulations, the constitutive relations for fluid-particle drag force are important. The exact filtered drag force $\overline{f_{gs,i}}$ can be obtained from fully resolved simulations via $\overline{f_{gs,i}}=\overline{\beta(u_{g,i}-u_{s,i})}$, where $\beta$ is the microscopic drag coefficient. It is well accepted in the literature \citep{Igci2011a,Ozel2013,Parmentier2012,Wang2010} that the coarse-grid resolved drag $\beta^{*}(\tilde{u}_{g,i}-\tilde{u}_{s,i})$ has a significant difference with $\overline{\beta(u_{g,i}-u_{s,i})}$ due to unresolved heterogeneous structures. The superscript '*' represents the value of the base function evaluated at the coarse grid using the filtered particle volume fraction $\bar{\phi}_s$ and the filtered phase slip velocities $\tilde{\bold{v}}$, e.g., $\beta^{*}=\beta|_{(\bar{\phi}_s,\tilde{\bold{v}})}$, $\left(\frac{\partial{\beta}}{\partial{\phi_s}}\right)^{*}=\left.\frac{\partial{\beta}}{\partial{\phi_s}}\right|_{(\bar{\phi}_s,\tilde{\bold{v}})}$ and so on. The approximation of $\beta$ could be obtained from a second order Taylor series expansion at $(\bar{\phi}_s,\tilde{\bold{v}})$ as follows
\begin{eqnarray}\label{eq13}
	\beta & \approx & \beta^{*}+\phi'_s\left(\frac{\partial{\beta}}{\partial{\phi_s}}\right)^{*}+(u''_{g,j}-u''_{s,j})\left(\frac{\partial{\beta}}{\partial{v_j}}\right)^{*}+\frac{{\phi'_s}^{2}}{2}\left(\frac{\partial{^{2}\beta}}{\partial{{\phi_s}^{2}}}\right)^{*}\nonumber\\
&& +\frac{(u''_{g,j}-u''_{s,j})(u''_{g,k}-u''_{s,k})}{2}\left(\frac{\partial{^{2}\beta}}{\partial{v_j}\partial{v_k}}\right)^{*}+\phi'_s(u''_{g,j}-u''_{s,j})\left(\frac{\partial{^{2}\beta}}{\partial{v_j}\partial{\phi_s}}\right)^{*} 
\end{eqnarray}
where $v_j=u_{g,j}-u_{s,j}$ is the slip velocity in $j$ direction. Then, the scaled filtered drag coefficient at $i$ direction, $\bar{\beta}_i=\overline{\beta|_{(\phi_s,\bold{v})}(u_{g,i}-u_{s,i})}/(\tilde{u}_{g,i}-\tilde{u}_{s,i})$, could be expressed as
\begin{eqnarray}\label{eq14}
	\bar{\beta}_i/\beta^{*} & \approx & 1+\underbrace{(a)\frac{\tilde{v}_{d,g,i}}{\tilde{v}_i}}_{(i)}+\underbrace{(b)\frac{\tilde{v}_{d,s,i}}{\tilde{v}_i}}_{(ii)}+\underbrace{(c)\frac{\tilde{v}_j\tilde{v}_{d,g,j}}{\tilde{v}^{2}}}_{(iii)}+\underbrace{(d)\frac{\tilde{v}_j\tilde{v}_{d,s,j}}{\tilde{v}^{2}}}_{(iv)}+\underbrace{(e)\frac{\tilde{v}_j\overline{v''_jv''_i}}{\tilde{v}^{2}\tilde{v}_i}}_{(v)}+\underbrace{(f)\frac{\overline{{\phi'_s}^{2}}}{(1-\bar{\phi}_s)\bar{\phi}_s}}_{(vi)}\nonumber\\
&& +\underbrace{(f)\frac{\overline{{\phi'_s}^{2}v''_i}}{(1-\bar{\phi}_s)\bar{\phi}_s\tilde{v}_i}}_{(vii)}+\underbrace{(g)\frac{\tilde{v}_j\overline{v''_jv''_i\phi'_s}}{\tilde{v}^{2}\tilde{v}_i(1-\bar{\phi}_s)}}_{(viii)}+\underbrace{(h)\frac{\tilde{v}_j\tilde{v}_k\overline{v''_jv''_k}}{\tilde{v}^{4}}}_{(ix)}+\underbrace{(h)\frac{\tilde{v}_j\tilde{v}_k\overline{v''_iv''_jv''_k}}{\tilde{v}^{4}\tilde{v}_i}}_{(x)}
\end{eqnarray}
\begin{equation}
\left. \begin{array}{l}
\displaystyle
(a) = [\bar{\phi}_s+\frac{(1-\bar{\phi}_s)\bar{\phi}_s}{\beta^{*}}]\left(\frac{\partial{\beta}}{\partial{\phi_s}}\right)^{*},\quad (b) = [(\bar{\phi}_s-1)+\frac{(1-\bar{\phi}_s)\bar{\phi}_s}{\beta^{*}}]\left(\frac{\partial{\beta}}{\partial{\phi_s}}\right)^{*}, \\[12pt]
\displaystyle
(c) = \frac{\bar{\phi}_s\tilde{v}}{\beta^{*}}\left(\frac{\partial{\beta}}{\partial{v}}\right)^{*}+\frac{(1-\bar{\phi}_s)\bar{\phi}_s\tilde{v}}{\beta^{*}}\left(\frac{\partial{^{2}\beta}}{\partial{v}\partial{\phi_s}}\right)^{*},  \\[12pt]
\displaystyle
(d) = -\frac{(1-\bar{\phi}_s)\tilde{v}}{\beta^{*}}\left(\frac{\partial{\beta}}{\partial{v}}\right)^{*}+\frac{(1-\bar{\phi}_s)\bar{\phi}_s\tilde{v}}{\beta^{*}}\left(\frac{\partial{^{2}\beta}}{\partial{v}\partial{\phi_s}}\right)^{*},\\[12pt]
\displaystyle
(e)=\frac{\tilde{v}}{\beta^{*}}\left(\frac{\partial{\beta}}{\partial{v}}\right)^{*},  \quad (f)=\frac{1}{2}\frac{(1-\bar{\phi}_s)\bar{\phi}_s}{\beta^{*}}\left(\frac{\partial{^{2}\beta}}{\partial{\phi_s^{2}}}\right)^{*},\\[12pt]
\displaystyle
(g)=\frac{(1-\bar{\phi}_s)\tilde{v}}{\beta^{*}}\left(\frac{\partial{^{2}\beta}}{\partial{v}\partial{\phi_s}}\right)^{*},\quad (h)=\frac{1}{2}\frac{\tilde{v}^{2}}{\beta^{*}}\left(\frac{\partial{^{2}\beta}}{\partial{v}^{2}}\right)^{*}.
\end{array} \right\}
\end{equation}
where $\tilde{v}_i=\tilde{u}_{g,i}-\tilde{u}_{s,i}$, $v''_i=v_i-\tilde{v}_i$ and $\tilde{v}=\sqrt{\tilde{v}_j\tilde{v}_j}$. Note that the subscript $\mathit{i}$ is a free index denoting the direction $\mathit{i}$, while $\mathit{j}$ and $\mathit{k}$ are summation indices.  Eq. (\ref{eq14}) is the approximated linearized expression of the normalized drag coefficient. This expression contains the gas drift velocity $\tilde{v}_{d,g,i}$, the solid drift velocity $\tilde{v}_{d,s,i}$, the scalar variance of the solid volume fraction $\overline{{\phi'_s}^{2}}$, turbulent-kinetic-energy-like terms (e.g. $\overline{v'_iv''_j}$, $\overline{v''_jv''_k}$) and some third-order fluctuating terms. The importance of all the terms in Eq. (\ref{eq14}) will be evaluated in the next section.

\section{Statistical analyses of filtered drag force}
In this section, we first evaluate the importance of each term in the proposed expression of the filtered drag force (\ref{eq14}) via theoretical computation of their coefficients based on  \citet{WenandYu}'s drag law in a wide range of solid volume fraction (from 0 to 0.6) and particle Reynolds number (from 0 to 100). We then validate the new expression against the most recent numerical results by \citet{Ozel2017} in low-Reynolds-number flows. 

\begin{figure}
  \centerline{\includegraphics[scale=1.0]{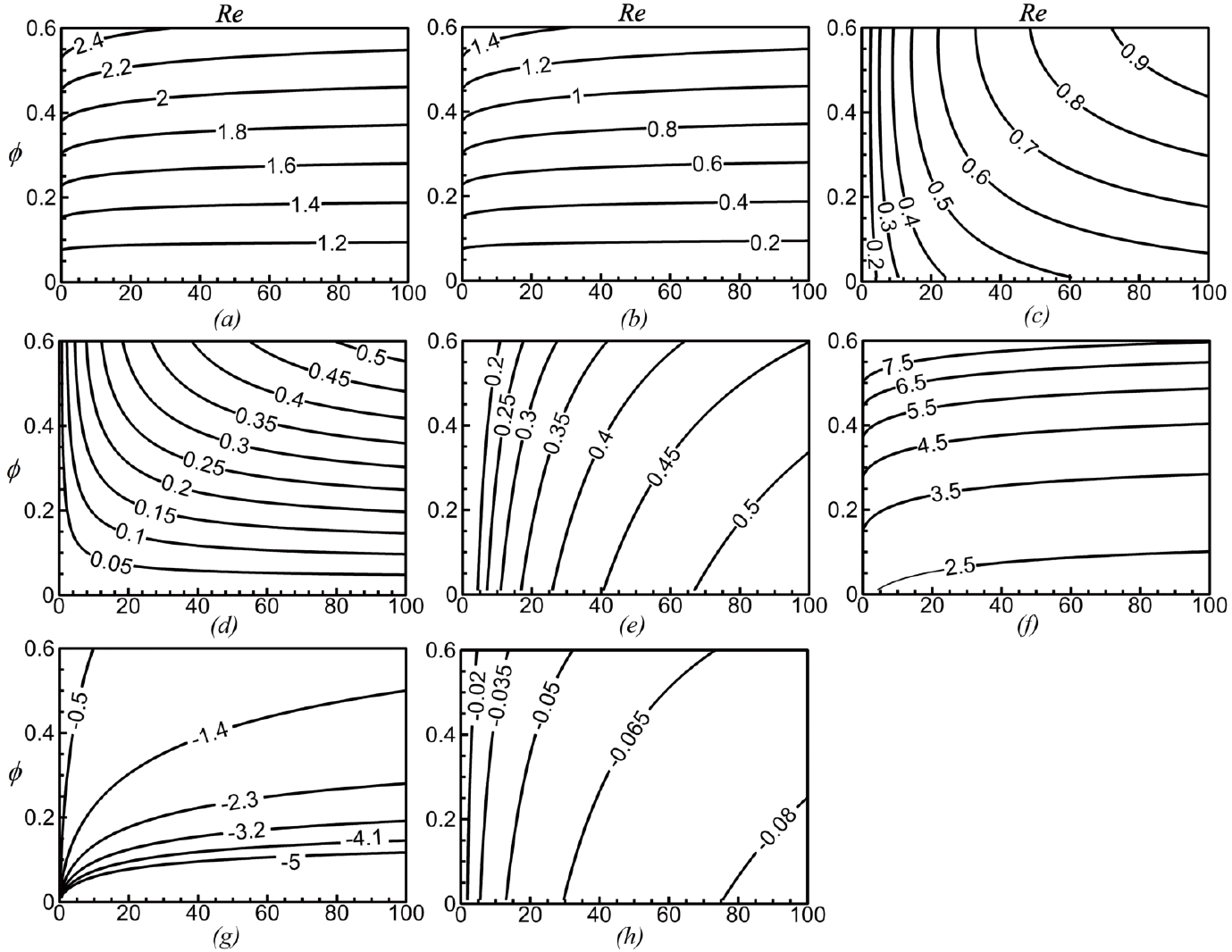}}
  \caption{The variation of coefficients in Eq. (\ref{eq14}) using \citet{WenandYu}’s drag law at different solid volume fractions and particle Reynolds numbers; the subtitle of each figure represents the coefficient marker in Eq. (\ref{eq14})}
\label{fig1}
\end{figure}

Each term in Eq. (\ref{eq14}) has been normalized with the filtered slip velocity and the filtered volume fraction, making the value of each term excluding their coefficients close to or less than 1.0. Thus, the coefficients could give a good estimation of the significance of the terms. The coefficients of each term in Eq. (\ref{eq14}) at different solid volume fractions and particle Reynolds numbers are depicted in figure \ref{fig1}. It reveals that the gas drift velocity term $(i)$ and the scalar variance of solid volume fraction term $(vi)$ have appreciable coefficients ($(a)$ and $(f)$) at large ranges of solid volume fraction and particle Reynolds number. This supports previous findings based on correlative analyses of numerical results\citep{Ozel2017,Rubinstein2017} in which the gas drift velocity and the scalar variance of solid volume fraction are found to be the two main markers in estimating the filtered drag force. It is noted that the coefficient $(f)$ is inherently positive and hence the scalar variance of solid volume fraction term $(vi)$ tends to increase the filtered drag force in any occasions. Considering that the gas is prone to flow through the dilute channels between clusters, the gas velocity $u_g$ unveils a positive correlation with $\phi_g$. Referring to Eq. (\ref{eq10}), we could conclude that the gas drift velocity is mostly negative and contributes to drag decrease at the filtered scale. This conclusion is consistent with numerous numerical simulations (e.g., \citet{Ozel2017,Schnei2017}). 

Additionally, though $\overline{{\phi'_s}^{2}v''_i}$ is a third order fluctuating quantity which is not mentioned in previous research, term $(vii)$ may not be omitted since the coefficient $(f)$ is non-negligible. The sign of $\overline{{\phi'_s}^{2}v''_i}$ mostly depends on the sign of the Favre fluctuating slip velocity, $v''$, in the very dense area and also in the very dilute area since only at these two occasions ${\phi'_s}^{2}$ attains significant values. $v''$ represents the difference between the local slip velocity $v_i$ and the Favre averaged slip velocity $\tilde{v}_i$ at the filtered scale. $v''$ vanishes at the locations where $v_i=\tilde{v}_i$. Denoting the solid volume fraction of such locations as $\phi_0$, it is known that, in general, $v''$ attains negative values at locations where $\phi_s>\phi_0$ and positive values at locations where $\phi_s<\phi_0$ since $v_i$ decreases in dense area and increases in dilute area. However, the change of $v_i$ with $\phi_s$ is not linear and is more profound in the dense area because the drag force has significant nonlinear increase with the increase of $\phi_s$\citep{Van2005Lattice}. Hence, ${\phi'_s}^{2}v''_i$ in the dense area dominates, making $\overline{{\phi'_s}^{2}v''_i}$ attains negative in most cases. This suggests that the third-order moment term functions like the gas drift velocity term, tending to decrease the filtered drag force.

The coefficient $(b)$ also attains positive values and the values increase rapidly as the solid volume fraction increases. This suggests that the solid drift velocity term $(ii)$ may become quite significant at moderate and high solid volume fractions. To coarsely estimate the values of the solid drift velocity, we consider the magnitude of Favre fluctuating slip velocity, $\bold{u}''_g-\bold{u}''_s$. It, if not less than, would be approximately of the same order of the magnitude of the slip velocity at the filtered scale level, which is $\tilde{\bold{u}}_g-\tilde{\bold{u}}_s$. Through the definition of the particle Reynolds number we have $\left|\tilde{\bold{u}}_g-\tilde{\bold{u}}_s\right|=\mu Re/\rho d$, where $\mu$ and $\rho$ are the dynamic viscosity, the density of the fluid phase and $d$ is the diameter of the particles. Clearly, $\left|\tilde{\bold{u}}_g-\tilde{\bold{u}}_s\right|$ vanishes at the low $Re$ limit. Hence $\bold{u}''_g \approx \bold{u}''_s$ holds approximately at low Reynolds numbers. Indeed, $\bold{u}''_g \approx \bold{u}''_s$ has been used by \citet{Christine1997} in estimating the time-averaged drag force. Later, \citet{Schnei2017} points out that this conclusion does not always hold in fluidized beds, which may be caused by the fact that his simualtions are not restricted in low-Reynolds-number regimes. Nevertheless, it is believed that $\bold{u}''_g$ and $\bold{u}''_s$, if not very close to each other, at least have the same sign in most occasions. Referring to Eqs. (\ref{eq10}) and (\ref{eq11}), also considering $\phi'_s=-\phi'_g$, we conclude that the solid drift velocity attains mostly positive values. Thus, since the coefficient $(b)$ is positive, the solid drift velocity term $(ii)$ is prone to increase the drag force at the filtered scale. Considering the relative magnitudes of these coefficients in low-Reynolds-number regime, Eq. (\ref{eq14}) could be simplified as
\begin{equation}\label{eq15}
	\bar{\beta}_i/\beta^{*}=1+(i)+(ii)+(vi)+(vii)
\end{equation}

In moderate- and high-Reynolds-number regimes, the increase of the Reynolds number leads to obvious increases in other coefficients, such as $(c)$, $(d)$, $(e)$, $(g)$ and $(h)$. This suggests that all terms in Eq. (\ref{eq14}) may have non-negligible impact on the filtered drag coefficient. However, the discussion on this is out of scope of the present study.

\begin{figure}
  \centerline{\includegraphics[scale=1.0]{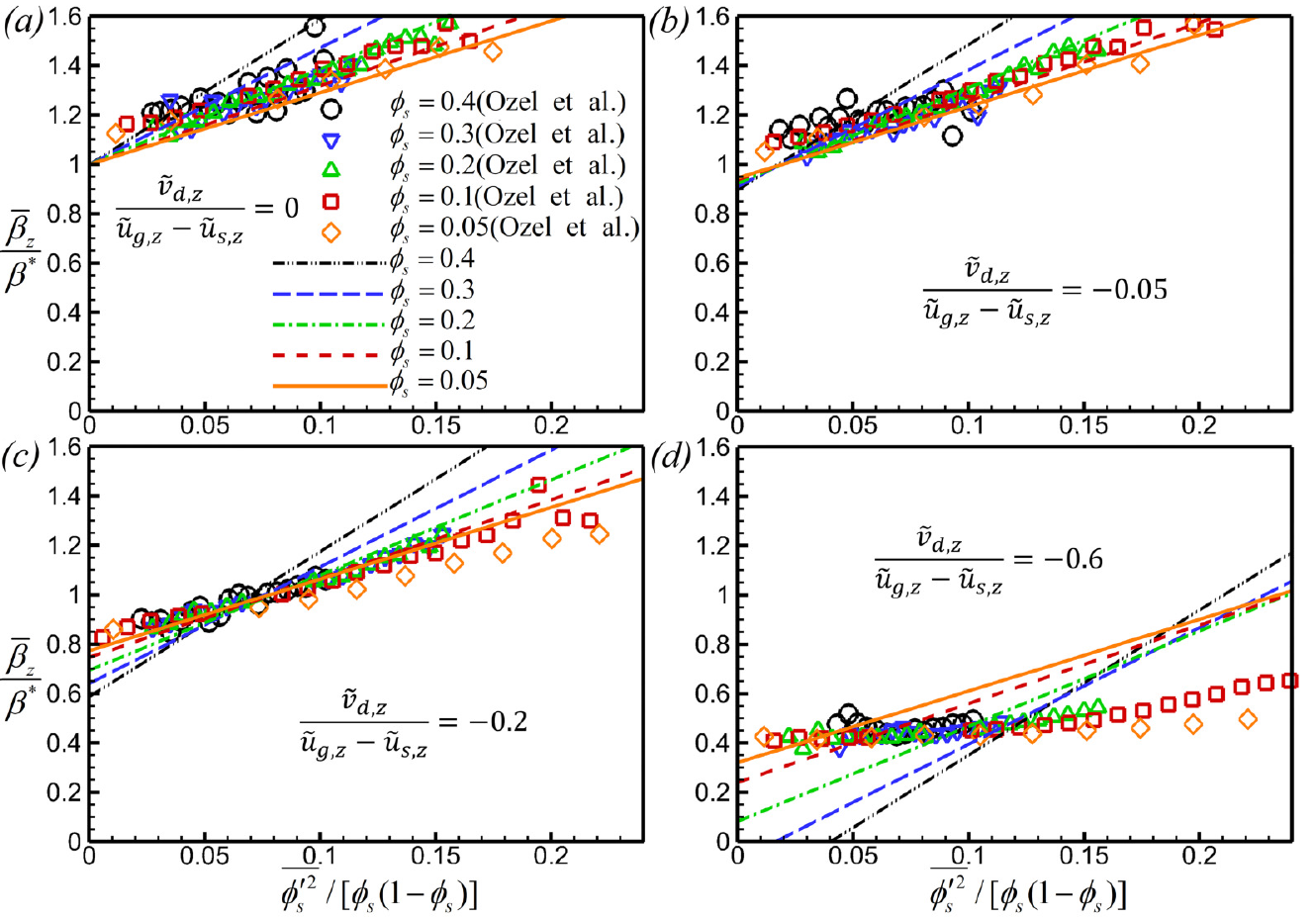}}
  \caption{Scaled filtered Eulerian drag coefficient as a function of the scalar variance of solid volume fraction for different filtered solid volume fractions at specific scaled drift velocities. The discrete symbols represent the numerical data from \citet{Ozel2017}, and the lines only account for three terms in Eq. (\ref{eq15}), which are, the constant term 1, gas drift velocity term $(i)$ and the term of the scalar variance of solid volume fraction $(vi)$.}
\label{fig2}
\end{figure}

The results in figure \ref{fig2} reveal the relationship between the scaled filtered drag coefficient and scalar variance of solid volume fraction $\overline{{\phi'_s}^{2}}$ at different gas drift velocities. It should be mentioned that the discrete numerical data provided is the overall scaled filtered drag coefficient $\bar{\beta}_z$ at the global flow direction $z$ obtained by \citet{Ozel2017}, while the lines represent our approximation results obtained using three terms in Eq. (\ref{eq15}), which are, the constant term 1, gas drift velocity term $(i)$ and the term of the scalar variance of solid volume fraction $(vi)$. Here we simply use the global Reynolds numbers of Ozel et al.'s data to estimate the local slip velocity in the approximation since the velocity distributions are not available. The agreements between the approximation results and these simulation cases are favorable at small solid volume fractions with small gas drift velocities since the solid drift velocity terms $(ii)$ and the third-order term $(vii)$ could be neglected in that situation (see figures \ref{fig2}(a) and \ref{fig2}(b)). The third-order moment $\overline{{\phi'_s}^{2}v''_i}$ in $(vii)$ is a higher correlation of $\overline{{\phi'_s}^{2}}$. Hence, it could be neglected at low $\overline{{\phi'_s}^{2}}$ and may have appreciable values at large $\overline{{\phi'_s}^{2}}$. With this in mind, the solid drift velocity term $(ii)$ is the only possible term responsible for the deviations between numerical data and the approximation results in the low $\overline{{\phi'_s}^{2}}$ range in figures \ref{fig2}(c) and \ref{fig2}(d). Hence, adding the solid drift velocity term is expected to diminish the deviations by increasing the filtered drag force in the low $\overline{{\phi'_s}^{2}}$ range. This validates the drag increase effect of solid drift velocity term $(ii)$ proposed previously in this paper and also demonstrates a negative correlation between gas and solid drift velocities.

Furthermore, considering that the third-order moment $\overline{{\phi'_s}^{2}v''_i}$ could be important at high $\overline{{\phi'_s}^{2}}$ and the solid drift velocity term $(ii)$ only attenuates the drag reduction, the obvious over-prediction of our approximation results at high $\overline{{\phi'_s}^{2}}$ in all subfigures in figure \ref{fig2} should be primarily caused by the third-order moment term $(vii)$. Thus, adding the third-order moment term into the computed lines is expected to diminish this over-prediction by decreasing the filtered drag force at high $\overline{{\phi'_s}^{2}}$. Close observation shows that the over-prediction of our approximation results becomes even profound as the gas drift velocity decreases. This implies that the third-order term is in general negative and has a positive correlation with the gas drift velocity since their coefficients $(a)$ and $(f)$ in Eq.(\ref{eq14}) are both inherently positive. The implication of negative values for $\overline{{\phi'_s}^{2}v''_i}$ is in line with the theoretical analysis made above. $\overline{{\phi'_s}^{2}v''_i}$ might also be dependent on the solid volume fraction since different degrees of over-prediction are observed in figure \ref{fig2} for different solid volume fractions.

\begin{figure}
  \centerline{\includegraphics[scale=1.0]{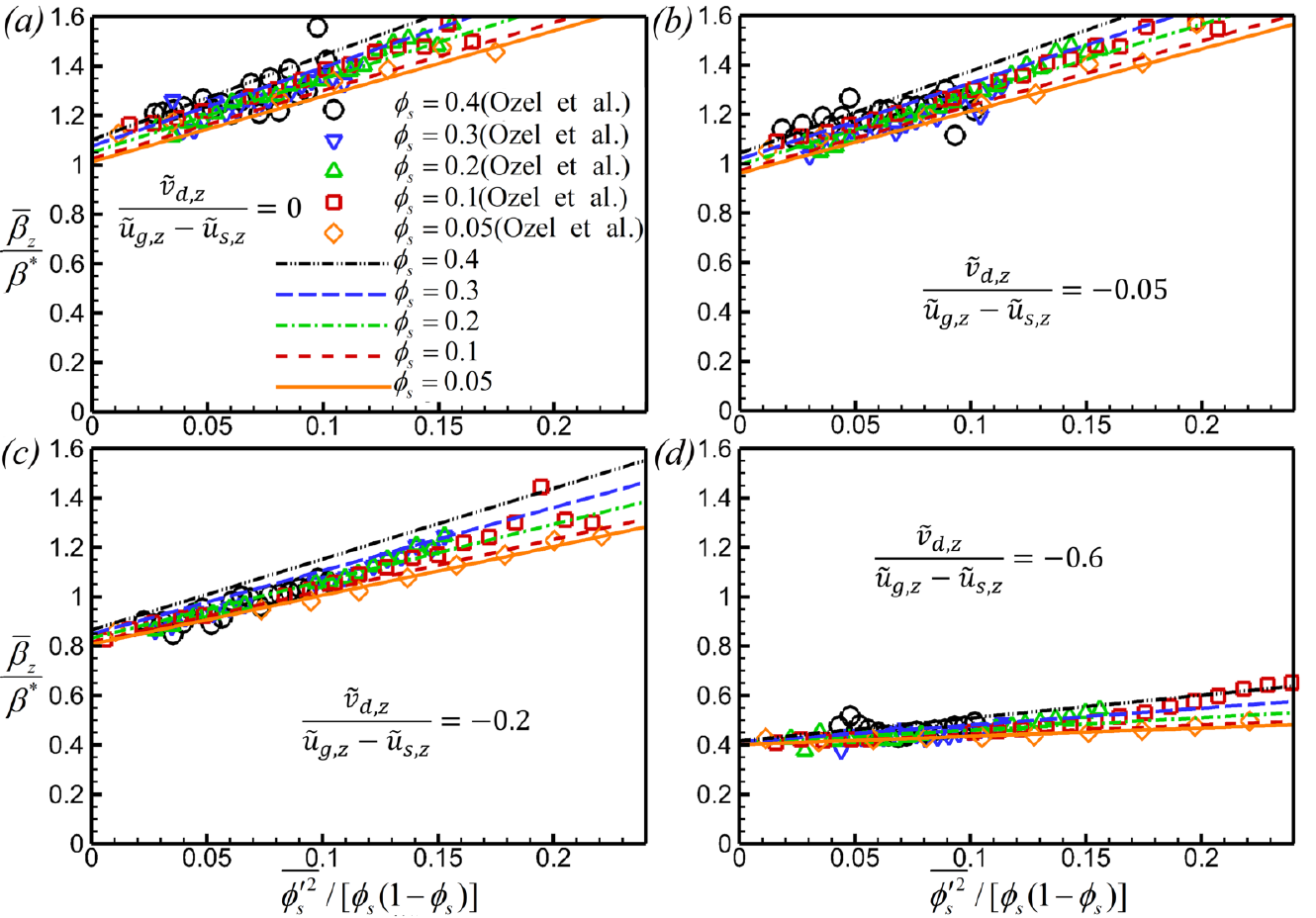}}
  \caption{Scaled filtered Eulerian drag coefficient as a function of the scalar variance of solid volume fraction for different filtered solid volume fractions at specific scaled drift velocities. The discrete symbols represent the numerical data from \citet{Ozel2017}, and the lines represent the Eq. (\ref{eq15}) using algebraic models (\ref{eq16}) and (\ref{eq17}).}
\label{fig3}
\end{figure}

Based on the analyses mentioned above, we tentatively propose the algebraic models for the solid drift velocity $\tilde{v}_{d,s,i}$ and the third-order moment $\overline{{\phi'_s}^{2}v''_i}$ as follows:
\begin{equation}\label{eq16}
	\frac{\tilde{v}_{d,s,i}}{\tilde{v}_i}=-\frac{\tilde{v}_{d,g,i}}{\tilde{v}_i}+0.1(1+\frac{\tilde{v}_{d,g,i}}{\tilde{v}_i})^{2}
\end{equation}
\begin{equation}\label{eq17}
	\frac{\overline{{\phi'_s}^{2}v''_i}}{\tilde{v}_i}=\overline{{\phi'_s}^{2}}\left \{ \frac{\tilde{v}_{d,g,i}}{\tilde{v}_i}-(1+\frac{\tilde{v}_{d,g,i}}{\tilde{v}_i})[(1+\frac{\tilde{v}_{d,g,i}}{\tilde{v}_i})\bar{\phi}_s+2(\frac{\tilde{v}_{d,g,i}}{\tilde{v}_i})^{2}(1-\bar{\phi}_s)] \right \}
\end{equation}

Note that the above two equations are only valid in the low-Reynolds-number regime, and Eq. (\ref{eq16}) is consistent with \citet{Christine1997}’s assumption especially for large gas drift velocities. Figure \ref{fig3} presents the comparison between \citet{Ozel2017}’s data and our proposed approximation Eq. (\ref{eq15}) with the models of (\ref{eq16}) and (\ref{eq17}). Favorable agreements are achieved for various gas drift velocities, demonstrating that the algebraic models (\ref{eq16}) and (\ref{eq17}) are suitable for prediction of the filtered drag force in the low-Reynolds-number regime. The agreements also indicate that the solid drift velocity term $(ii)$ and the third-order moment term $(vii)$ are non-ignorable even at low Reynolds numbers. This does not contradict with the finding in the literature that the gas drift velocity and scalar variance of solid volume fraction are two main markers in estimating the scaled filtered drag coefficient (e.g., see \citet{Rubinstein2017,Ozel2017}). This is simply because the effects of term $(ii)$ and $(vii)$ could be absorbed into the two markers when only the correlations between the filtered drag force and the two markers are concerned. Nevertheless, the comparison indicates that the four sub-grid quantities in Eq. (\ref{eq15}) are key parameters controlling the behavior of the filtered drag force. The respective impact of them on the filtered drag force revealed in the comparison is in line with that obtained through the theoretical analysis reported above, which is, the gas drift velocity and the third-order moment are responsible for the drag reduction relative to the microscopic drag force, whereas the solid drift velocity and scalar variance of solid volume fraction tends to attenuate this reduction.

\section{Summary}
We have performed a Taylor series expansion of a microscopic drag coefficient $\beta$, and obtained a more accurate expression of filtered drag force by reserving the second order terms. Our results imply that besides the two main markers recognized in the literature, i.e., gas drift velocity $\tilde{v}_{d,g,i}$  and scalar variance of solid volume fraction $\overline{{\phi'_s}^{2}}$, the solid drift velocity $\tilde{v}_{d,s,i}$  and a third-order moment $\overline{{\phi'_s}^{2}v''_i}$  are also significant for obtaining an accurate prediction of the filtered drag force even at low Reynolds numbers.

The mechanism of how the four sub-quantities affect the filtered drag force is illustrated via the theoretical analysis for the first time. The gas drift velocity and the third-order moment are found to be responsible for the drag reduction at the filtered scale, whereas the solid drift velocity and scalar variance of solid volume fraction tends to attenuate this reduction. It is also found that the solid drift velocity is prone to give more significant impact at moderate and high solid volume fractions because its coefficient rapidly increases with the solid volume fraction. The third-order moment clearly manifest itself only when the $\overline{{\phi'_s}^{2}}$ is relatively large and this surely happens as shown in figures \ref{fig2}(c) and \ref{fig2}(d) since pronounced inhomogenieties prevail in those situations.

The numerical results from \citet{Ozel2017} are used to help understand the mechanism. Apparent deviations exist between the numerical results and the filtered drag correlation that only accounts for the effects of the gas drift velocity term and the term of scalar variance of solid volume fraction. The addition of the terms of the solid drift velocity and the third-order moment gives favorable agreement with the numerical results. The values of the solid drift velocity and the third-order moment are unavailable in this study. We estimate them as functions of the gas drift velocity and scalar variance of solid volume fraction through expressions (\ref{eq16}) and (\ref{eq17}). This indeed establishes algebraic models for predicting the solid drift velocity and the third-order moment. However, more numerical simulations should be performed to further validate our proposed approximation and to formulate and refine closures for the four sub-grid quantities based on obtained quantities in the coarse-grid simulations. Furthermore, the characteristics of the filtered drag force with different microscopic drag laws, e.g., BVK model\citep{Beetstra2007}, \citet{Tang2015}'s model, \citet{Tenneti2011}'s model and Zhou \& Fan’s model\citep{Zhou2015a,Zhou2015b} also should be assessed systematically to explore the effect of the choice of the microscopic drag law in the future.

\section*{Ackonwledgements}
We are grateful to the financial support by the National Natural Science Foundation of China (91634114, 11602190), the Natural Science Foundation of Jiangsu Province for Youth (BK20160390) and the China Postdoctoral Science Foundation (2017M613129).

\bibliographystyle{jfm}
\bibliography{Arxive}

\end{document}